\pdfoutput=1
\documentclass{JINST}
\usepackage{graphicx}
\usepackage{epsfig}
\usepackage{epstopdf}
\usepackage{amssymb}

\title{\bf Absolute Calibration of a Large-diameter Light Source}

\author{{J.T. Brack$^a$\thanks{Corresponding author.}, R. Cope$^{a,b}$,
  A. Dorofeev$^a$, B. Gookin$^a$, J.L.~Harton$^a$, 
  Y.~Petrov$^a$,~A.C.~Rovero$^c$\thanks{Member of Carrera del Investigador, CONICET}}\\
\llap{$^a$}Colorado~State~University,~Department~of~Physics,~Fort~Collins~CO~80523,~USA,\\ 
\llap{$^b$}Now at Cornell~University,~Ithaca,~NY,\\
\llap{$^c$}Instituto de Astronom{\'\i}a y F{\'\i}sica del Espacio
     (CONICET-UBA), Buenos Aires, Argentina\\
E-mail: \email{brack@colostate.edu}\\
}

\abstract{
A method of absolute calibration for large aperture optical systems is
presented, using the example of the Pierre Auger Observatory fluorescence
detectors.  A 2.5~m diameter light source illuminated by an
ultra--violet light emitting diode is calibrated with an overall uncertainty of
2.1~\% at a wavelength of 365~nm.}


\begin{document}

\section{Introduction}
\label{sec:Intro}

The Pierre Auger Observatory has been designed to study the origin and the
nature of ultra high-energy cosmic rays, which have energies above
$10^{18}$~eV~\cite{EA}.  The Observatory, located in Malarg\"ue,
Argentina, consists of two detector systems which provide
independent information on Extensive Air Showers (EAS) initiated by cosmic ray
interaction in the Earth's atmosphere.  The Surface Detector (SD) is composed of
over 1660 water Cherenkov detectors located on a triangular array of 1.5~km
spacing covering an area of 3000~km$^{2}$ measuring EAS secondary particles
reaching ground level~\cite{SD}. The Fluorescence Detector (FD) consists of 27
telescopes distributed in buildings on the periphery of the SD overlooking the
array~\cite{NIM}\cite{HEAT}.  

The UV-nitrogen fluorescence light produced along the core of the particle
shower in the air is registered by the FD on clear nights, while the SD operates
continuously.  Events observed by both FD and SD provide the link from the FD,
which is absolutely calibrated, to provide the energy calibration
~\cite{SDcal} of the higher statistics data from the SD.  The
reconstruction of air shower longitudinal profiles and the determination of the
total energy of reconstructed showers depends on the conversion of ADC counts
from the FD to an absolute light flux at the aperture for each pixel.  This
conversion must be available for each observing night and for all wavelengths of
the FD response range. To achieve this objective three different FD calibration
procedures are performed~\cite{rob}: relative,
multi--wavelength~\cite{multiwave}\cite{lasercalib}, and absolute calibrations.
The measurement of the intensity of the light source used for the absolute
calibration is discussed in this paper.

The calibrated 2.5~m diameter portable light source (referred to as the ``drum''
because of its appearance) is used at the FD apertures, providing uniform
illumination to each pixel at a single wavelength~\cite{NIM}.  The known flux from the
light source and the response of the acquisition system give the required
calibration for each pixel.  The procedure we used previously to calibrate the
drum at the laboratory has been outlined elsewhere~\cite{jeff}\cite{pune}.  The
drum and electronics are designed to have a pulsed intensity approximately
matching the intensity of a typical EAS at the FD. This light flux is too low to
be measured directly by a photodiode in the lab by about a factor of 10,000, and
too bright to be calibrated using PMTs and single-photon counting techniques.
Our new procedure described here uses the $1/{r^2}$ attenuation of a point
source with distance to allow absolute calibration of a small light source at a
short distance using the calibrated photodiode and to relate this intensity to
that of the drum at large distance using a PMT.

In Secs.~\ref{sec:technique} and~\ref{sec:Lim_Valid} we introduce the technique
and its limitations due to finite sizes of the source and detectors.  The new
light source and controlling electronics are described in
Sec.~\ref{sec:hardware}.  Details of the new procedure in the laboratory and
results are presented in Sec.~\ref{sec:drumcal}, and a discussion of
uncertainties is given in Sec.~\ref{sec:syst_Uncertain}.

\section{Technique: Use of $1/{r^2}$ attenuation of light}
\label{sec:technique}

The factor of about 10,000 needed to accommodate the difference in intensities
required by the calibrated photodiode and the intensity level of the drum is achieved,
conceptually, by the following steps: The drum light source is pulsed, and at
some large distance (in practice about 15 m in the lab in Malargue) a PMT is
used to record a histogram of the intensity.  Next, a small-diameter light
source (which we call the rail light source) is set up at this same distance,
and the intensity is adjusted to give the same PMT histogram intensity as for
the drum.  Then, without changing its intensity, the rail light source is moved
to an optical bench in a dark box where the pulsed absolute intensity can be
measured directly using a calibrated photodiode at a short distance (in practice
about 10~cm).

The PMT is used only as a measure of the drum intensity relative to that of the
rail light source, so no  PMT calibration is required. The same  LEDs and driving
electronics are used for the drum and the rail light source, ensuring identical
pulse characteristics.

In practice, several conditions and systematic checks are required to ensure
reliability of the measurement.  First, since the angle of incidence for a
photon on the front glass PMT face and photocathode affects PMT response, the
distance between the PMT and the drum light source must be
sufficiently large that the drum provides light normally incident on the PMT  (see
Sec.~\ref{sec:Lim_Valid}).  The 15~m distance limits the angle of incidence
to less than 5 deg.

A potentially large source of error is reflection from the walls of the dark
hall or from the interior of the optical bench dark box.  Light blocking baffles were used in
both environments.  To confirm $1/r^2$ behavior, the rail  light source
was moved along a rail in 1~m steps from about 10~m to 15~m distance from  the PMT, 
and in 10~cm steps from about 1~m to 10~cm from the NIST-calibrated photodiode in the dark
box.  Stray light and reflections will in general not vary as $1/r^2$, so this is an
important test.

\section{Limits of $1/{r^2}$ validity for extended sources and detectors}
\label{sec:Lim_Valid}

The number of photons $N_{D}$
observed by a detector of radius $R_{D}$ emitted from a Lambertian disk of radius $R_{E}$
can be written:
\begin{equation}
\label{eq:finite_size_detect}
N_D=\frac{N_E}{2}\times\left( \
	\left( 1 + \frac{r^2+R_D^2}{R_E^2} \right) - \
	\sqrt{\left( 1 + \frac{r^2+R_D^2}{R_E^2} \right)^2 - 4 \frac{R_D^2}{R_E^2}}
	\right)
\end{equation}
where $N_E$ is a total number of photons emitted by the disk, and $r$ is the
distance between emitter and the detector.
Equation~\ref{eq:finite_size_detect} assumes that the planes of the detector and
emitter are perpendicular to the optical axis passing through the central points. 

\begin{figure}
\begin{center}
\includegraphics[width=.95\textwidth]{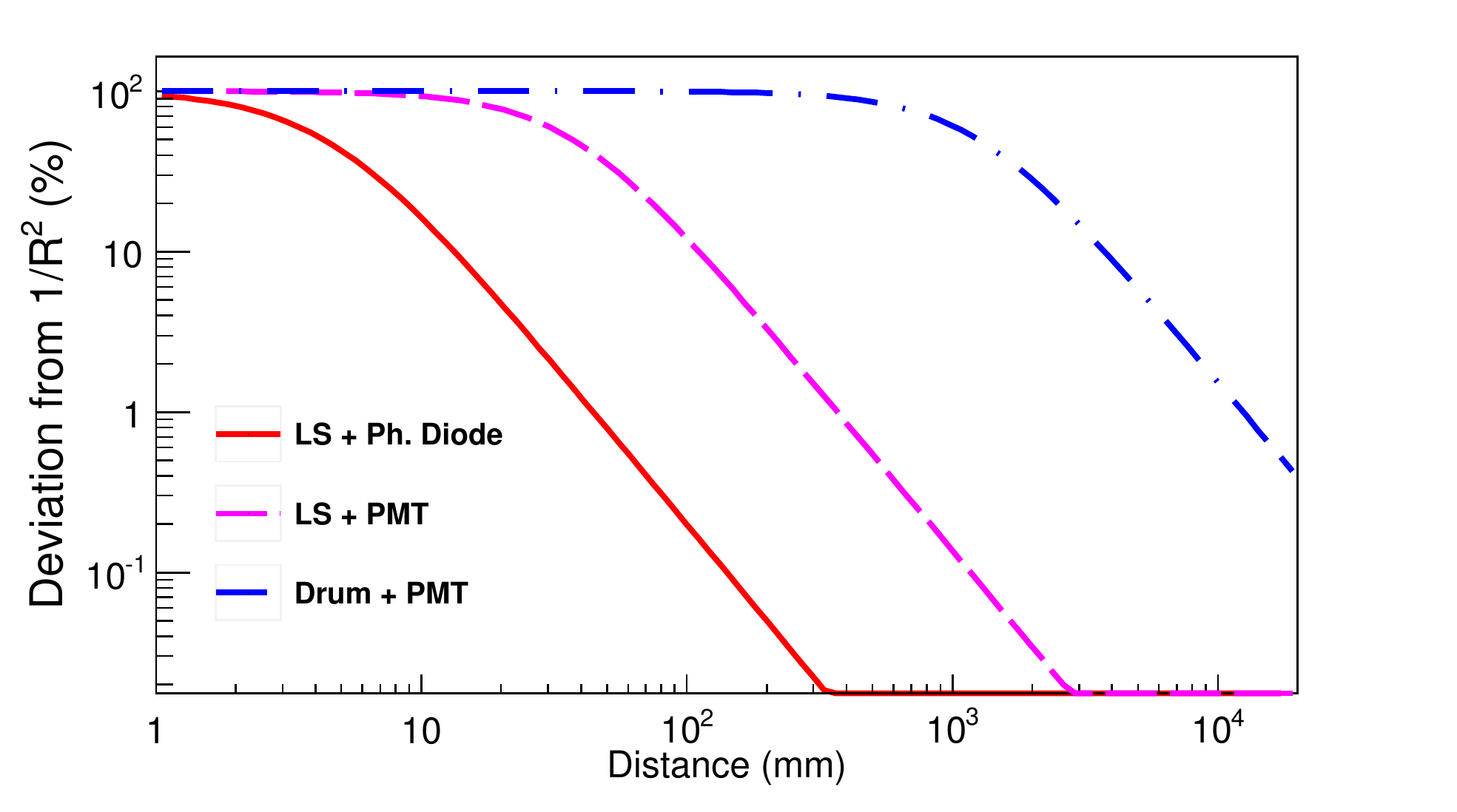}
\caption{Deviation from $1/{r^2}$ behavior as a function of distance 
between source and detector for the sizes (radii) of the sources involved in the
procedure: drum (1.25~m), PMT (37~mm), and Light Source (2~mm).}
\label{fig:deviations}
\end{center}
\end{figure}

Two limits of Eqn.~\ref{eq:finite_size_detect} are of interest here. 
For $r=0$ we obtain the following approximations:
\begin{equation}
\label{eq:finite_size_detect_risclose}
N_D= N_E \times \left\{ \begin{array}{ll} 
	1 &\mbox{ if $R_D \geq R_E$} \\
	\frac{R_D^2}{R_E^2} &\mbox{ if $R_D < R_E$ }
	\end{array} \right.
\end{equation}
indicating that when the emitter is smaller than the detector all emitted
photons are detected, and that if the emitter becomes larger than the detector
the fraction of detected photons is proportional to the ratio of the areas.  
At the limit of large distances, $r \gg \sqrt{R_D^2 + R_E^2}$,
Eqn.~\ref{eq:finite_size_detect} reduces to a description of a finite detector
observing a point source:
\begin{equation}
\label{eq:finite_size_detect_risfar}
N_D= \frac{N_E \cdot R_D^2}{r^2}.
\end{equation}
Figure~\ref{fig:deviations} shows the percentage deviation from
$1/{r^2}$ behavior for detector-source combinations corresponding to our
equipment.  The solid red line represents Eqn.~\ref{eq:finite_size_detect} evaluated
for the geometry used in the dark box, where the radii of the photodiode and the
rail light source are 4~mm and 2~mm, respectively, and the deviations from
point-like $1/{r^2}$ behavior vary from less than 1\% to 0.05\% at distances
between 100 and 1000~mm, as used in the dark box measurements.  Similarly, the
dashed magenta line is evaluated for the PMT--rail light source geometry, for which
the radii are 37~mm and 2~mm, and deviations are below 1\% at distances of 1~m,
while the smallest distance using this combination in our measurements is 10~m
for which approximately 0.01\% deviation is calculated.  The dot-dash blue line, for the
PMT-drum combination, uses radii of 37~mm and 1250~mm.  At a distance of 15~m,
as used in the dark hall, the deviation is again less than 1\%.  For all distances
and geometry used in the calibration described here, deviations are well below
1\%.  (See Fig.~\ref{fig:Darkbox_Residual} for actual distances used.)

Radiant intensity of the emitter $I$ in units of $\left[\frac{photons}{sr}\right]$ 
can be calculated from the  number of emitted photons, $N_E$, using
Eqn.~\ref{eq:drum_radiant_intensity}: 
\begin{equation}
\label{eq:drum_radiant_intensity}
I(\Theta) = \frac{N_E}{\pi}\cdot \cos(\Theta) = I_0 \cdot \cos(\Theta)
\end{equation}
where $\Theta$ is the angle from the normal to the surface of the emitter. Our
measurement of the drum intensity results in $I_0$.

\section{Hardware}
\label{sec:hardware}
\subsection{Drum}
\label{sec:Drum_Desc}
\begin{figure}
\begin{center}
\includegraphics[width=.8\textwidth]{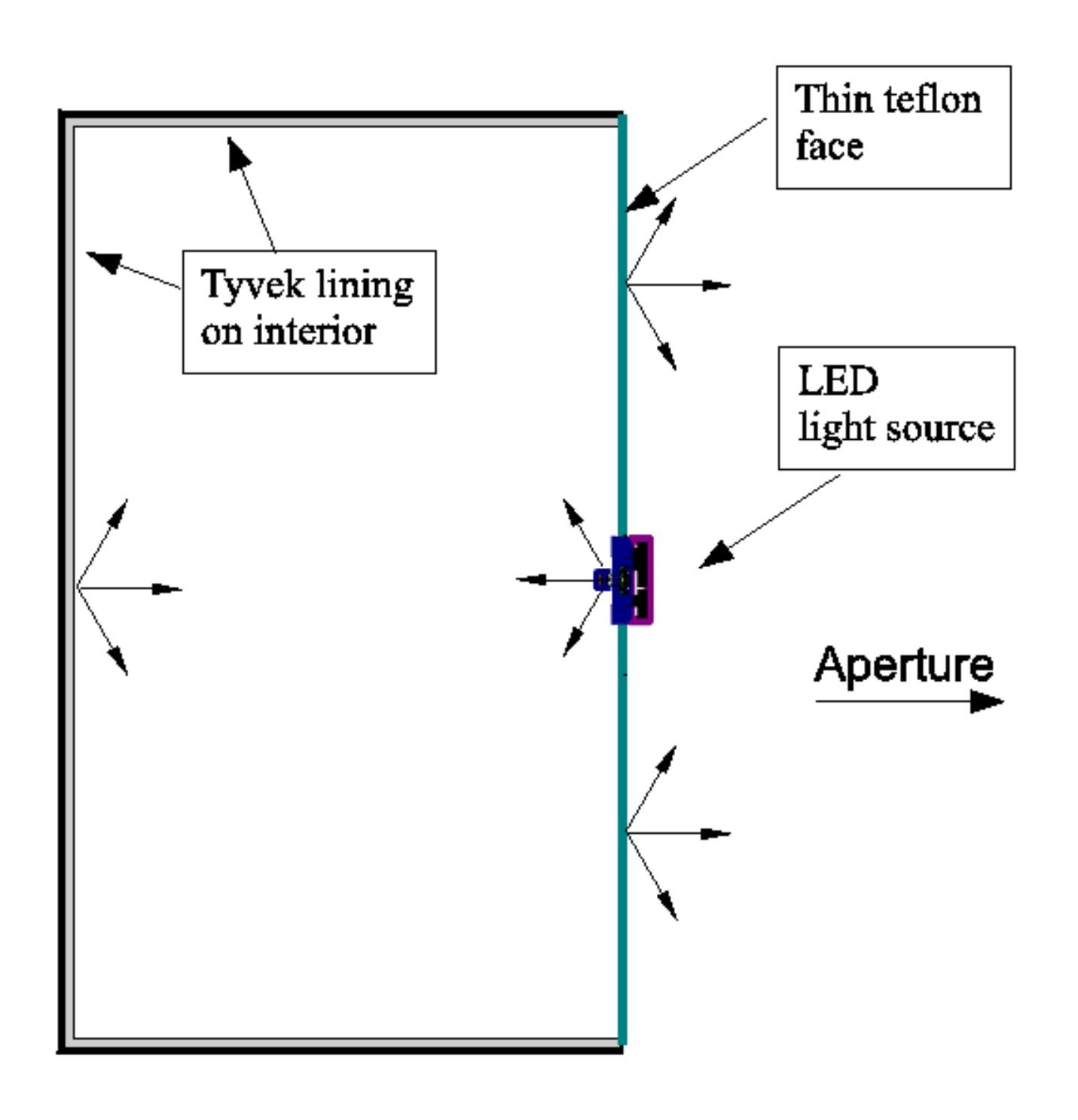}
\caption{Schematic of the calibration light source and drum.  The drum is
  1.4~m deep and 2.5~m in diameter.}
\label{fig:drum}
\end{center}
\end{figure}

The portable light source (drum) has been designed to uniformly illuminate all
440 pixels in a single camera simultaneously. The drum is a cylinder of 2.5 m
diameter and 1.4~m deep constructed in sections, using laminations of paper
honeycomb core and aluminum sheet (see Fig.~\ref{fig:drum}). The sides and back
surfaces of the drum interior are lined with Tyvek$^{\circledR }$, a material
diffusively reflective in the UV. The front face of the drum is a 0.38~mm thick
Teflon$^{\circledR }$ sheet, which transmits light diffusively.  We have
developed a stabilized UV light source (see Sec.~\ref{sec:lightsource}) that is
placed on the front drum surface illuminating the interior so that the light
experiences diffusive reflection from the Tyvek before being diffusively
transmitted by the Teflon front surface of the drum.  The multiple diffuse
reflections result in a more uniform and Lambertian light source (see
Sec.~\ref{sec:drumtests}).  This same light source is used on the rail with a
different diffuser.

The FD apertures are 2.2~m in diameter, thus part of the 2.5~m diameter drum
face is masked during use for FD calibration.  During absolute calibration of
the drum in the laboratory, described below, a 2.2~m diameter opaque mask is
mounted on the front face of the drum, mimicking the aperture condition.

\subsection{LED light source}
\label{sec:lightsource}

The LED and associated components are contained in an aluminum box
(18$\times$15$\times$4.5~cm), which is mounted directly on the face of the drum
illuminating the interior.  When configured as the rail light source, the same
electronics box is mounted on the rail in the dark hall at varying distances
from the PMT for calibration measurements. Different diffusers are used in the
two configurations.

The LED is mounted on a Peltier device with a radiator and fan for temperature control;
these components protrude from one side of the box.  To enhance uniform
distribution of photons, a Teflon diffuser covers the LED.  For use in the drum,
the diffuser is cylindrical.  Photons exit radially, making at least one
diffusive reflection from the Tyvek-lined reflector cup mounted inside the drum
before hitting the inner surfaces of the drum.  For use on the rail and in the dark box, this
cylindrical diffuser is replaced with a 2.5~cm diameter Tyvek disk mounted in a
2.5~cm diameter, 2.5~cm long pipe.  A 2.5~cm diameter black disk with a 4~mm diameter hole 
masks the Teflon at the exit of the pipe.

\begin{figure}[h]
\begin{center}
\includegraphics[width=.95\textwidth]{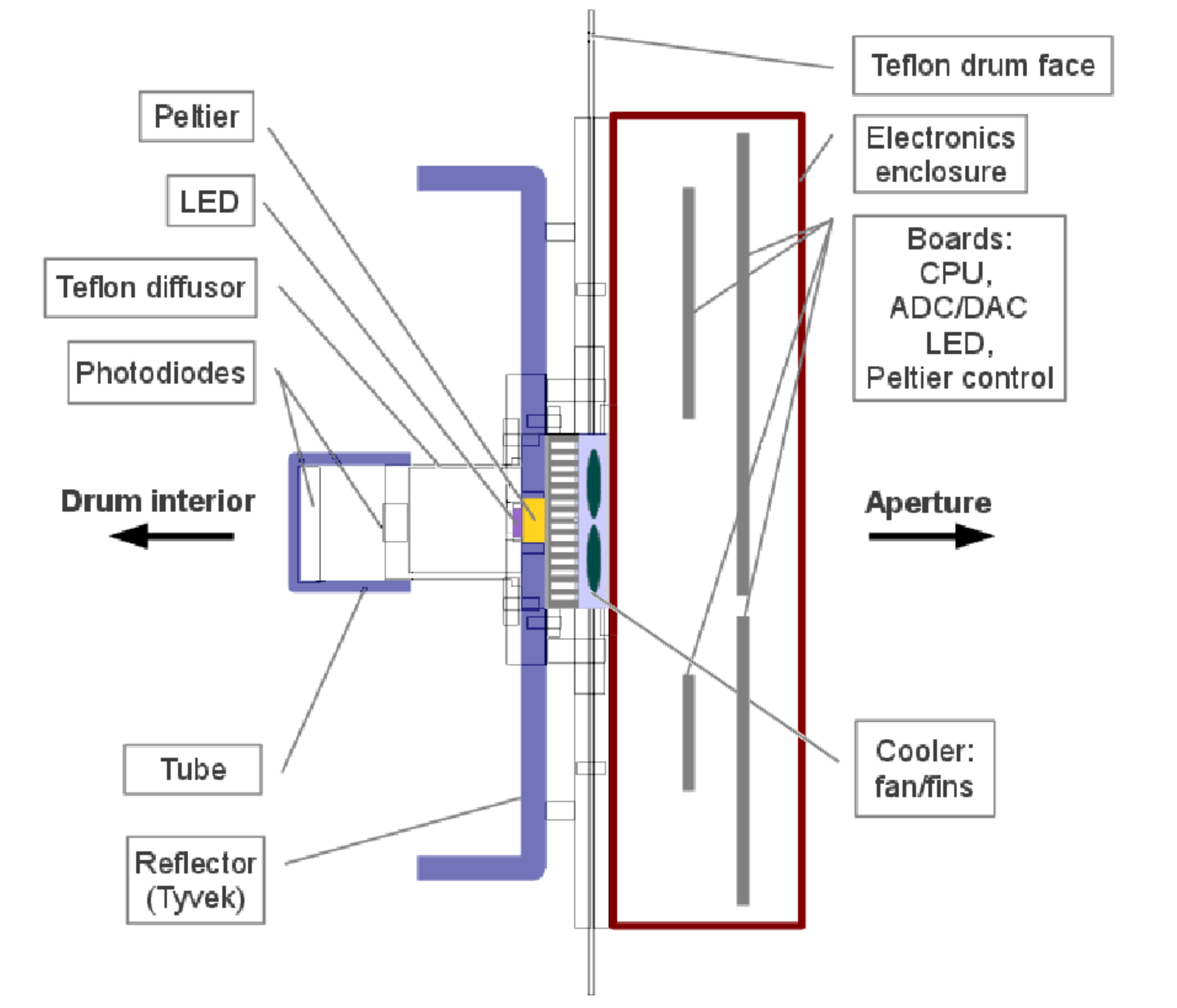}
\caption{Schematic diagram of the light source assembly, mounted on the drum,
  showing the position of the temperature controlled LED, the Teflon diffuser
  and the monitoring silicon detector.  The reflector cup is 15~cm in diameter,
  and is permanently mounted inside the drum.}
\label{fig:light-source}
\end{center}
\end{figure}

The light source, mounted in the reflector cup on the drum face, is shown in
Fig.~\ref{fig:light-source}, with the cylindrical diffuser in place.  

\subsection{Light source electronics}
\label{sec:electronics}

The electronics package includes a 600 MHz BlackFin BF537 processor running
$\mu$Clinux, controlling a Xilinx Spartan3A programmable logic device on a
separate board, which also holds four 12-bit DACs controlling LED current and
two 12-bit ADCs for reading the PMT and light source monitoring
photodiode signals.  The DACs and ADCs run at 100 MS/s.  The linux OS runs cgi
code controlling the light source pulse via specially designed web pages.  Pulse
shape, width, amplitude, and rate are programmable from the web page.
Traces from the ADCs are
stored in 32 Mb onboard memory and  are  read out via ethernet connection.

\begin{figure}[h]
\begin{center}
\includegraphics[width=.95\textwidth]{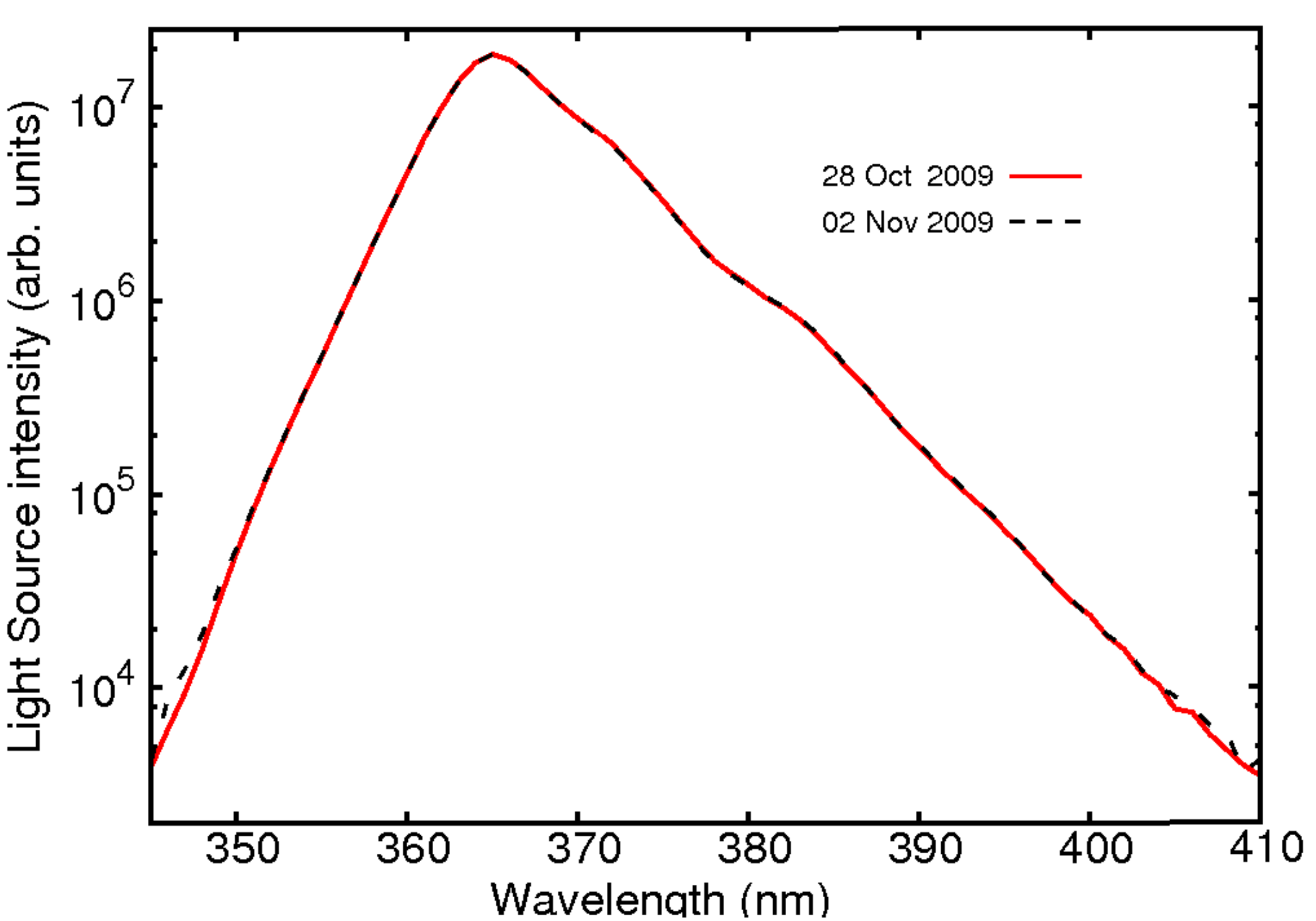}
\caption{Spectral stability of the LED in the lab. The spectra were taken five days
apart. The stability of total number of photons is better than 0.1\%.}
\label{fig:led365-stability}
\end{center}
\end{figure}

Stability of the LED~\cite{nichia} junction temperature is crucial for maintaining constant
optical output as a function of wavelength, and for stability over time
Fig.~\ref{fig:led365-stability} shows measurements of LED emission taken
several days apart from a continuously pulsing LED.  
Tests at Colorado State University of high output LEDs run
at large duty factors or high current have shown that the central wavelength can
shift by as much as 10 nm 
accompanied by an enhanced tail in the long wavelength end of the output
distribution, if insufficient temperature control is provided. To control
junction temperature as much as possible, the LED is mounted directly on the
surface of a Peltier device using an electrically non-conductive thermal paste. 
The Peltier is similarly
mounted on a radiator.  A dedicated microcontroller mounted on a separate board
in the electronics package controls the Peltier current with feedback from a
thermistor mounted at the LED.  Under the pulsed current
conditions used for our measurements, the controlled LED surface temperature is
stable at $22\pm0.5$C.  This temperature and the ambient temperature in the
electronics box are readout from the CAN bus.

\subsection{Uniformity of drum emission}
\label{sec:drumtests}

The drum light source has been constructed to provide a uniform flux of photons
through the FD apertures during calibration.  This is accomplished by using the
diffusively reflecting and transmitting materials described
in~\ref{sec:Drum_Desc}.  Validity of these Lambertian properties of the drum can
be found in~\cite{jeff} where the drum was viewed at angles from 0 to 25 degrees
with a CCD and the change in surface intensity was negligible.

\subsection{Calibration reference standard}
\label{sec:calrefstd}

The calibration reference is a UDT100 silicon photodiode~\cite{UDT}, calibrated
by NIST and
available as a calibration standard~\cite{NIST}, equipped with a 0.4999~cm$^2$
mask.  We use three such calibrated photodiodes for comparison of performance.  The
calibration of the photodiode at 365 nm is given by NIST as  0.119 A/W, with k=2
relative expanded uncertainty of 1.6\%, corresponding to a 1$\sigma$ uncertainty 
of ~0.8\%, provided the associated uncertainties in the NIST calibration measurements
are normally distributed.

\begin{figure}[h]
\begin{center}
\includegraphics[width=.45\textwidth]{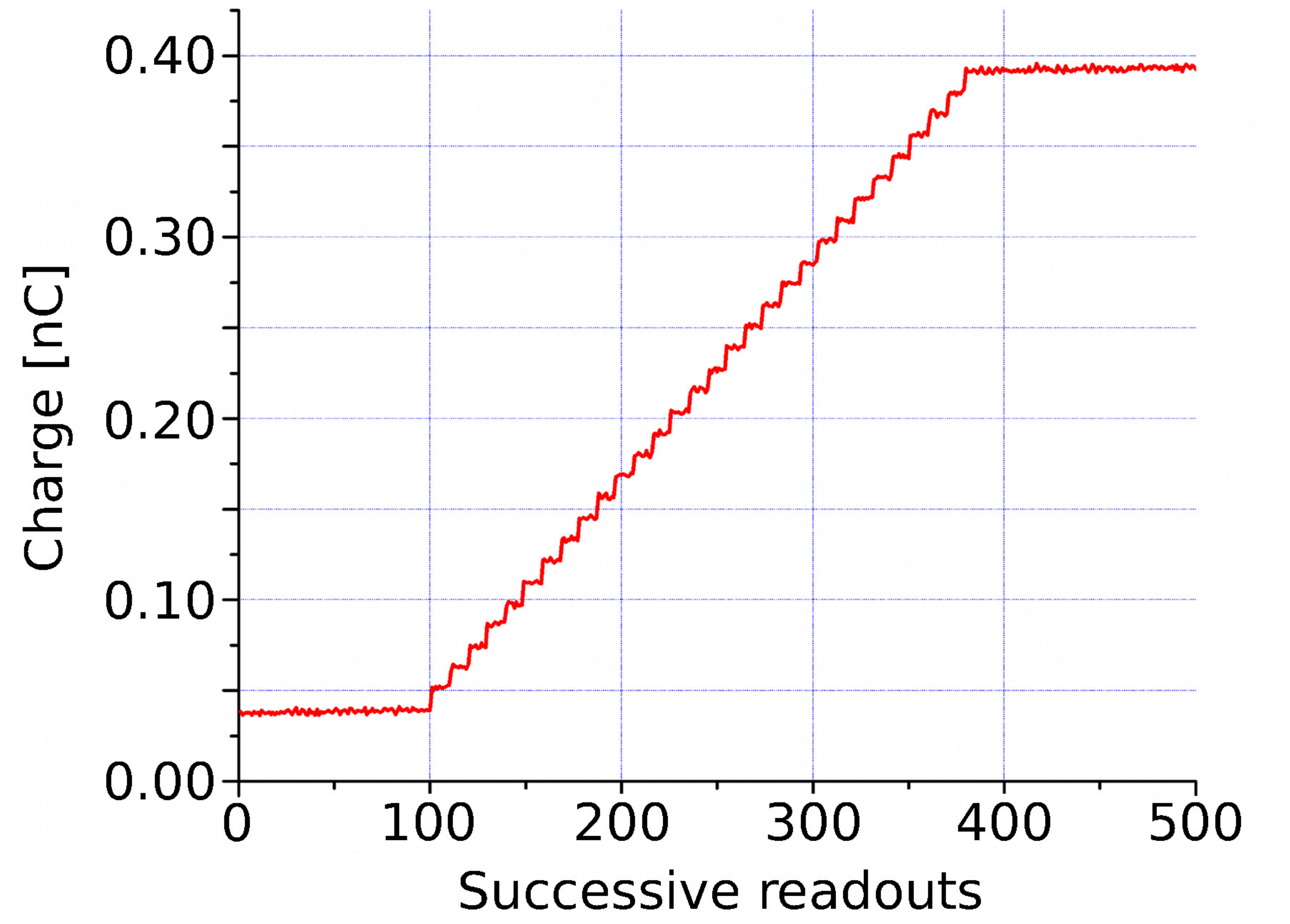}~~
\includegraphics[width=.45\textwidth]{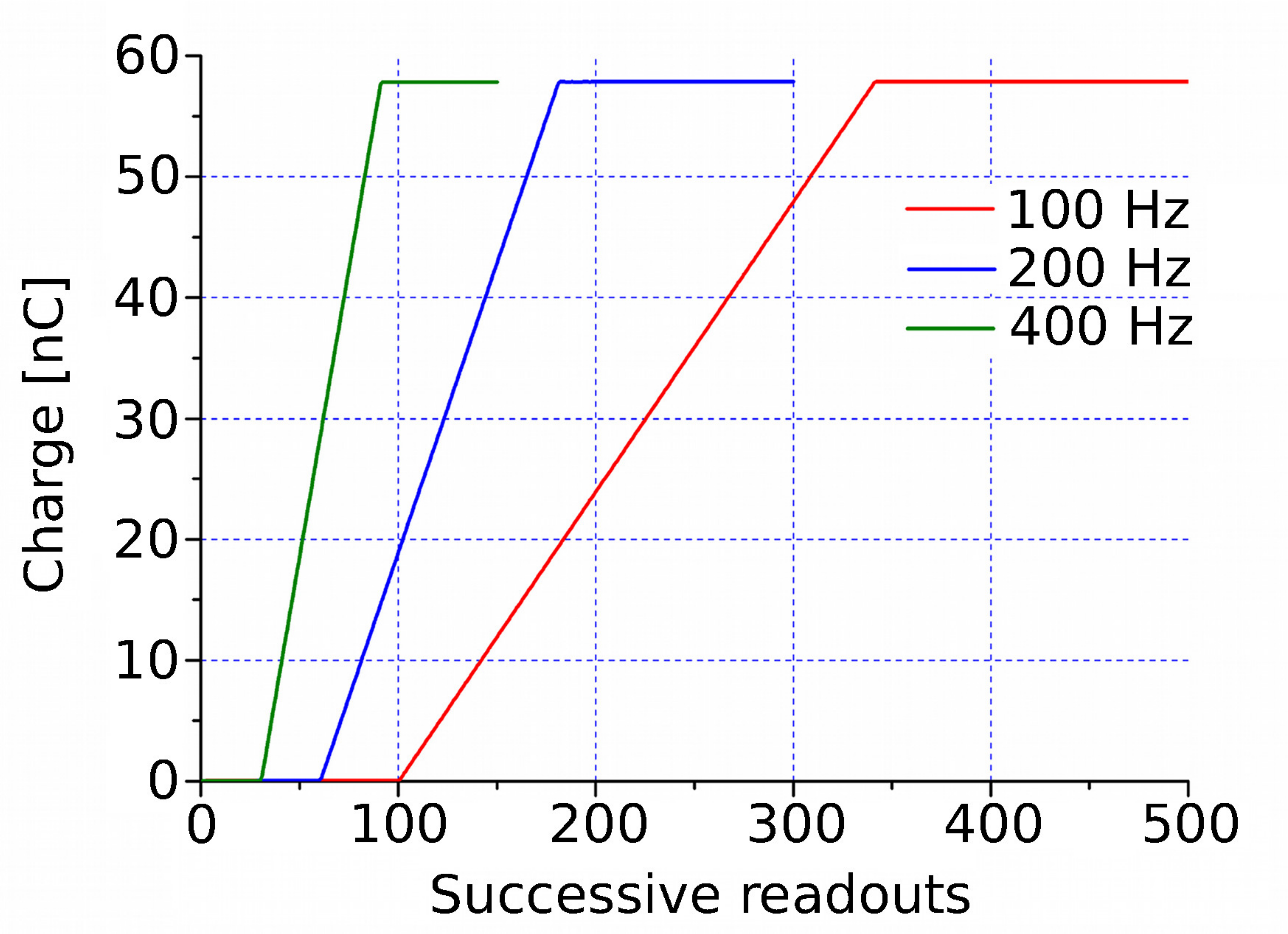}
\caption{Integration of the NIST-calibrated photodiode current using the
  Keithley 6514 electrometer in Q-mode. {\it Left:} 30 pulses of the LED at the rate
  of 1 Hz.  The sampling rate is faster than 1 Hz so one can see clear steps
  when each pulse happens and the accumulated charge increases. {\it Right:} 5000
  pulses submitted at different rates illustrate independence of the total
  collected charge from the pulsing rate.}
\label{fig:q-mode-readout}
\end{center}
\end{figure}

The photodiode is read out using a Keithley 6485 picoammeter.
Figure~\ref{fig:q-mode-readout} shows the output of the picoammeter in Q-mode.
Each time bin corresponds to an output cycle of the ammeter, corresponding to an
integration period.  Each step corresponds to a cycle during which the LED was
pulsed at a typical intensity.  The flat regions before and after the pulsing
show the rate of integration of the photodiode dark current, which is on the
order of 1 pA for these 1 cm$^2$ large-area photodiodes at room temperature.

\subsection{Dark Hall and Dark Box}

\begin{figure}[h]
\begin{center}
\includegraphics[width=.95\textwidth]{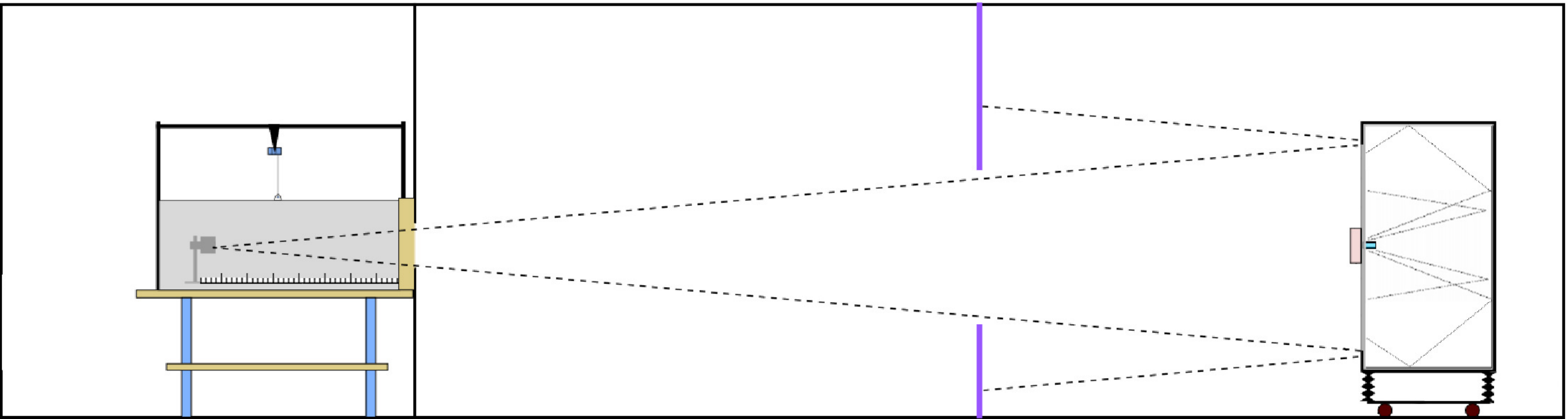}
\includegraphics[width=.95\textwidth]{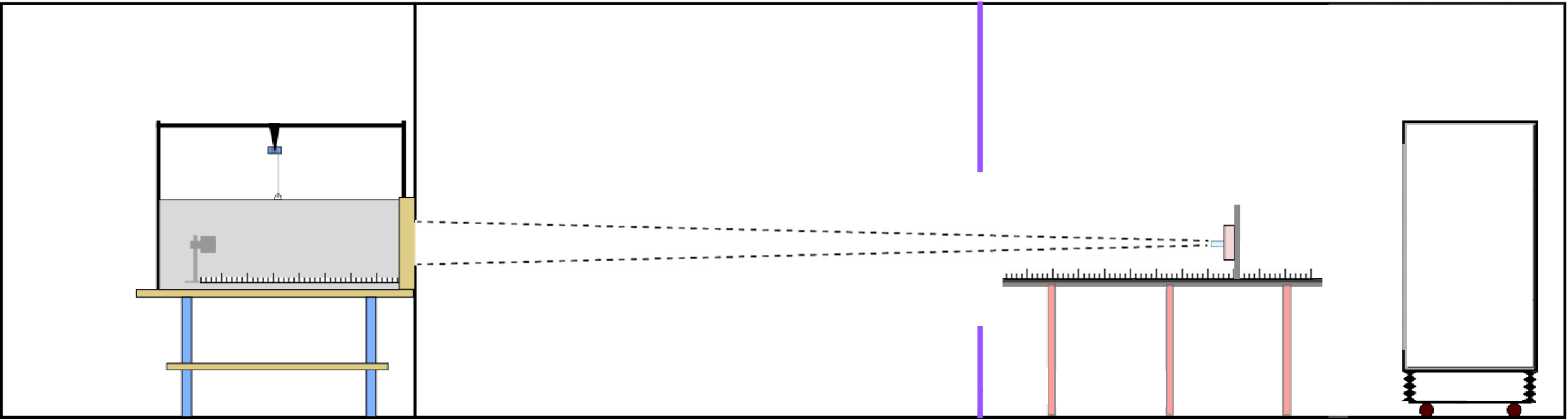}
\caption{A schematic of the calibration dark hall and equipment. {\it Top:} as set up
for drum intensity measurements. {\it Bottom:} as set up with the rail light
source for measurements of its intensity before transfer to the dark box with
the calibrated photodiode.}
\label{fig:yev-darkhall}
\end{center}
\end{figure}

The dimensions of the dark hall are $4\times4\times17$~m.  The layout is shown
in Fig.~\ref{fig:yev-darkhall}.  A 0.30~m diameter hole in the end wall allows the
addition of a $0.6\times0.6\times1.5$~m dark box in an adjoining room.  Curtains
at the mid-point of the dark hall create a light blocking baffle, preventing
primary reflections from the walls, ceiling and floor from reaching the entrance
to the dark box, which forms a second light
baffle.

\begin{figure}[h]
\begin{center}
\includegraphics[width=.5\textwidth]{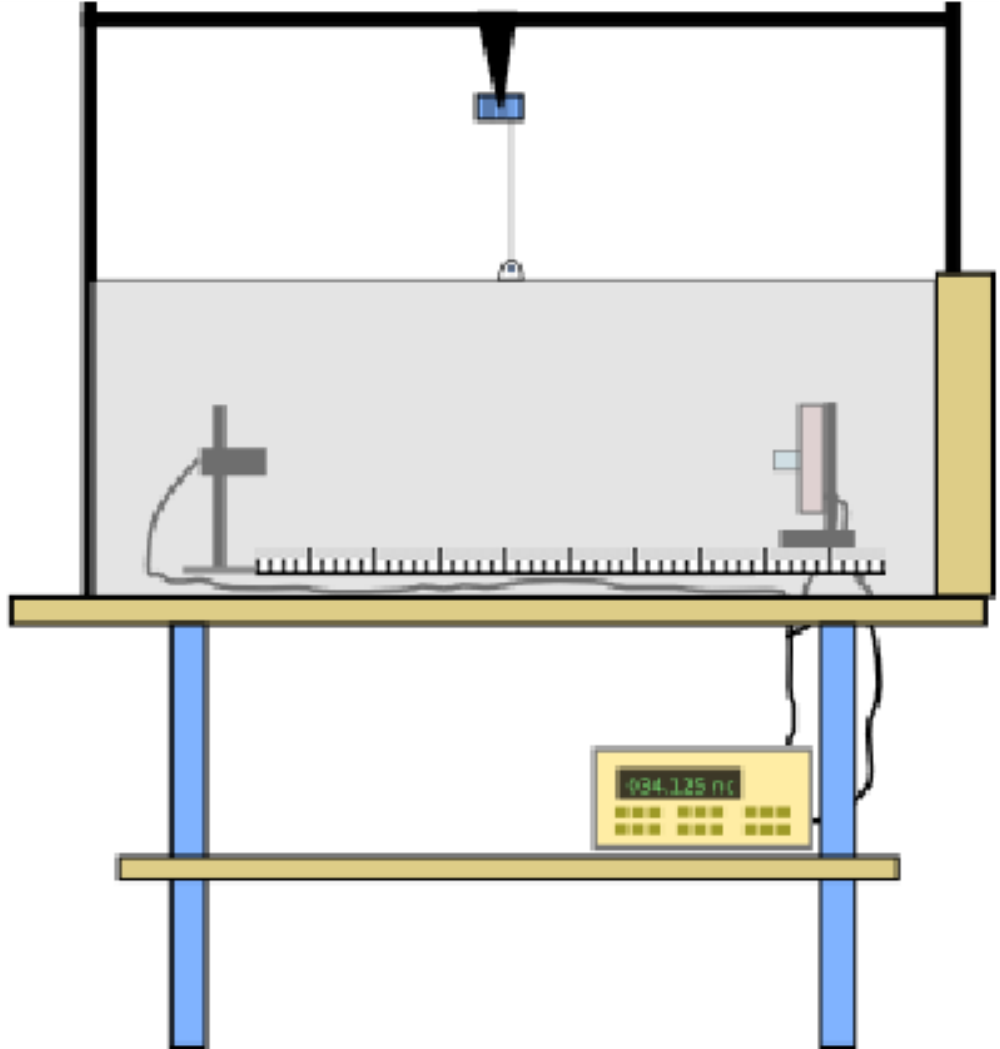}
\caption{Schematic of the calibration dark box with rail light source and calibrated
  photodiode.}
\label{fig:yev-darkbox}
\end{center}
\end{figure}

A cart mounted on a rail provides a movable mount for the rail light source,
which can be positioned at points from 10 to 15~m from the PMT face.

In the dark box, a linear actuator positions the rail light source at
points from 10~cm to 1~m from the calibration photodiode remotely, without
opening the box.  The centers of the drum, the rail light source, the PMT, and
the photodiode are on--axis with the two baffles.

\section{Drum absolute intensity calibration}
\label{sec:drumcal}
\begin{figure}[ht]
\begin{center}
  \includegraphics[width=.65\textwidth,angle=-90]{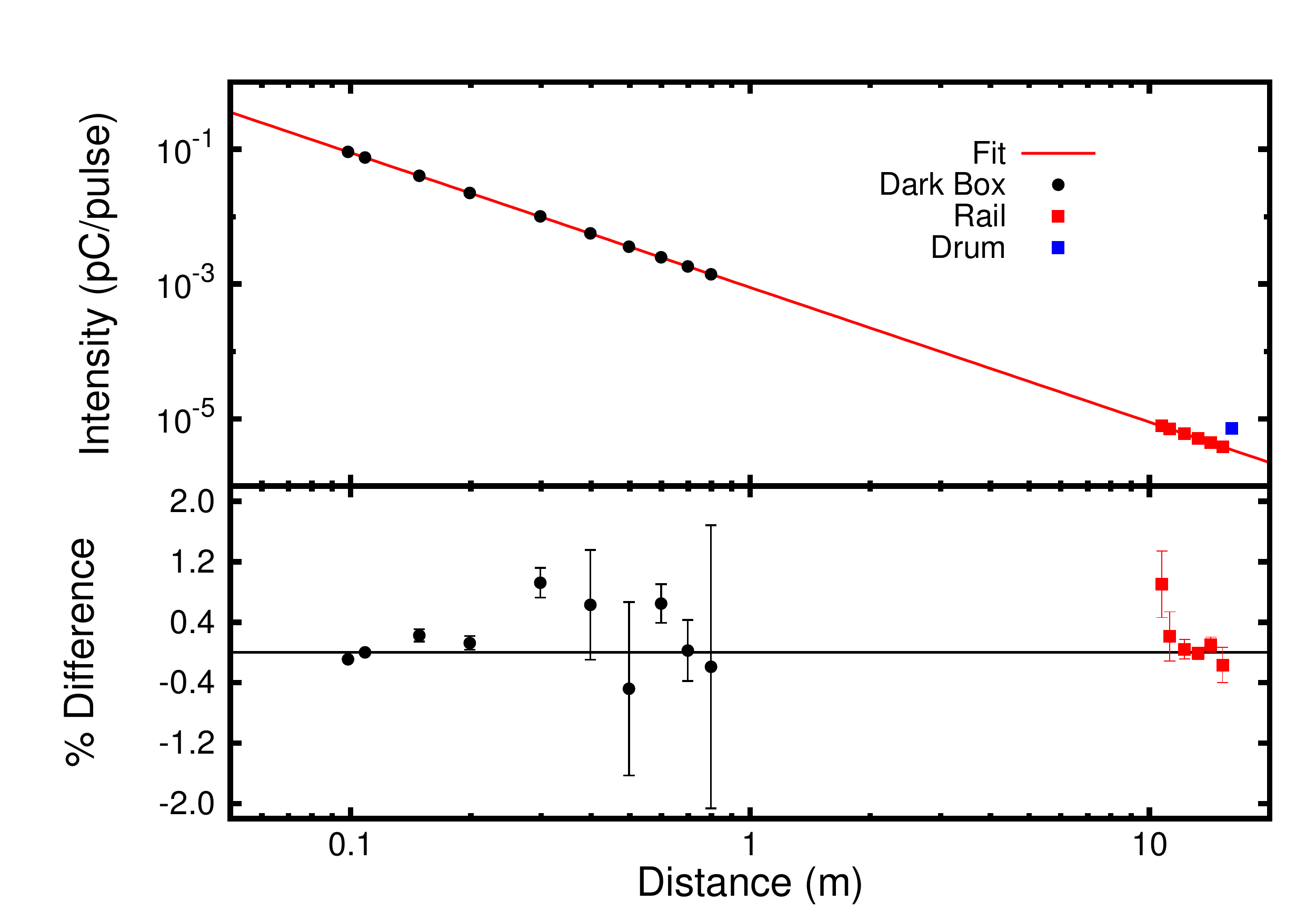}
  \caption{Measurements of PMT and photodiode responses to the light sources,
	and differences from a $1/{r^2}$ fit (red line).   The fit is made to only the
	black points, corresponding to measurements using the calibrated photodiode.
	Red points represent measurements using the PMT with the light source on
	the rail, and are normalized as a group to the fit.  The blue point is the
	PMT measurement of the drum intensity with the same PMT normalization factor applied.
	The residuals are shown on the bottom axis as percentage difference from the
	$1/{r^2}$ fit.}
 \label{fig:Darkbox_Residual}
\end{center}
\end{figure}

Following the technique described above, we positioned a PMT in the dark box at
a distance of $r_D=15.5$~m from the drum (see the upper diagram in
Fig.~\ref{fig:yev-darkhall}).
Then a series of PMT traces was recorded, each
corresponding to a  single  5~$\mu$s pulse, and the average of the ADC counts per pulse 
$H^{Drum}(r_D)=H^{Drum}$ was measured (see the blue square on Fig.~\ref{fig:Darkbox_Residual}).
Next, the light source was dismounted from the drum, the cylindrical Teflon
diffuser was removed, and the 2~mm diffuser was installed in its place -- thus
forming the rail light source.  The rail light source was then mounted on the
rail (see lower diagram in Fig.~\ref{fig:yev-darkhall}), and its intensity was
adjusted to give a PMT response similar to that observed with the drum.  A
series of PMT traces was recorded, each corresponding to a single 5~$\mu$s
pulse from the rail light source, and again the average per pulse was measured
$H^{Rail}(r_i)=H^{Rail}_i$, but now
at six different positions along the rail $r_i$ between 10.75 and 15.25~m from
the PMT; the corresponding points are marked as red squares on
Fig.~\ref{fig:Darkbox_Residual}.  
The rail light source was never placed at the full drum distance (the drum
was physically in the way) and it was purposely set at a slightly lower
intensity such that, as its position on the rail was varied, the response
spanned the measured drum intensity.  It is important that the PMT
detection system is linear over the range of responses, as demonstrated in
Fig. 8, and that the drum intensity is contained within the linear region.
In principle, the drum could have been much farther away, provided these
conditions were satisfied.

The rail light source was then moved into the dark box, and the PMT was replaced
by the calibrated photodiode.  The LED was flashed in the same manner as on the rail,
and the charge per pulse $Q(r_j)=Q_j$ was measured, now in absolute units of
$\left[\frac{C}{pulse}\right]$ with the photodiode dark current subtracted. We
did this at several distances $r_{j}$ between 10~cm and 1~m.  These measurements
are marked as black dots on Fig.~\ref{fig:Darkbox_Residual}.

In these measurements we have three combinations of light sources with
detectors: drum with PMT; rail light source with PMT; and rail light source with
calibrated photodiode.  The LED has a non-zero width in wavelength of emission, and
the diffusers can affect the spectrum from the drum and rail light
source. Additionally, the PMT and photodiode have different wavelength-dependent
responses.  The convolution of the spectra of these three sources with the
wavelength-dependent responses of the two detectors plays a role in the
extraction of the drum intensity.

We define the response of a detector at a particular wavelength $\lambda$ to be
$R_D(\lambda)=\frac{Signal_D}{n_D(\lambda)}$ where $Signal_D$ is the readout of
the detector exposed to some source (in ADC counts for the PMT; in Coulombs for
the calibrated photodiode) and $n_D(\lambda)$ is the number of photons as a function
of wavelength incident on a particular detector.  We write the total number of
photons incident on a detector over the source spectrum as $N_D$ in terms of
detector response integrated over the source spectrum
\begin{equation}
\label{eq:detected_photons_vs_signal}
N_D = \frac{Signal_D}{\int_\lambda{R_D(\lambda) Spectrum_E(\lambda) d\lambda}} = 
	\frac{Signal_D}{\Omega^E_D}
\end{equation}
where $Spectrum_E(\lambda)$ is the normalized emission spectrum of the source
$E$. Thus $\Omega^E_D$ is a constant for each source--detector combination which
depends on the wavelength-dependence of the source emitter and detector.  We
calculate these constants using our measurements of the 
optical properties of the diffusive properties
of the materials and the bare spectrum of the LED.

The drum radiant intensity $I^{Drum}_0$, in units of $\left[\frac{photons}{sr
    \times pulse}\right]$, can then be written using equations
Eqns.~\ref{eq:finite_size_detect_risfar},
	 \ref{eq:drum_radiant_intensity}, and \ref{eq:detected_photons_vs_signal} as
\begin{equation}
\label{eq:drum_intensity_nofph}
I^{Drum}_0 = \frac{N^{Drum}_E}{N^{Rail}_E}~I^{Rail}_0 = 
	\frac{H^{Drum} \cdot r^2_D}{<H^{Rail}_i \cdot r^2_i>} 
	\frac{\Omega^{Rail}_{PMT}}{\Omega^{Drum}_{PMT}} \times 
	\frac{<Q_j \cdot r^2_j>}{A \cdot \Omega^{Rail}_{Photodiode}}
\end{equation}
where $N^{Drum}_E$ and $N^{Rail}_E$ are the total number of photons emitted by
the drum and rail light sources, respectively, $I^{Rail}_0$ is the radiant
intensity of the rail light source, and $A$ is the area of the photodiode (mask
area and uncertainty provided by NIST).  The wavelength dependence is integrated
over in the three numbers $\Omega^{Rail}_{PMT}$, $\Omega^{Drum}_{PMT}$, and
$\Omega^{Rail}_{Photodiode}$.  As one can see from
Eqn.~\ref{eq:drum_intensity_nofph}, the detector responses appear as
$(Response)\cdot(Distance)^2$, which are constants for each source and detector
combination (assuming $1/r^2$ behavior), and therefore we take the averages of
these constants over the different distances for the rail light source and for
the single distance for the drum measurements.

The response of the NIST calibrated photodiode sets the absolute intensity scale 
in units of $\left[C\right]$ for all the PMT responses. A ratio of
the photodiode response to the PMT response 
in units of 
$\left[\frac{C}{ADC count}\right]$ can be written as
\begin{equation}
\label{eq:pmt_to_phd_ratio}
\frac{<Q_j \cdot r^2_j>}{<H^{Rail}_i \cdot r^2_i>}
\end{equation}
This conversion factor is then applied to all PMT measurements ($H^{Rail}_i$ and
$H^{Drum}$) individually.  The upper portion of Fig.~\ref{fig:Darkbox_Residual}
shows a graphical representation of this process.  Each point has
uncertainties $\frac{\sigma}{\sqrt{N}}$ on the mean of N repeated
measurements, and are too small to be seen on this plot.  

Eqn.~\ref{eq:drum_intensity_nofph} has four natural factors:
\begin{equation}
\label{eq:drum_intensity_divided}
F_1 = \frac{H^{Drum} \cdot r^2_D}{<H^{Rail}_i \cdot r^2_i>};
F_2 = \frac{\Omega^{Rail}_{PMT}}{\Omega^{Drum}_{PMT}};
F_3 = <Q_j \cdot r^2_j> ;
F_4 = \frac{1}{A \cdot \Omega^{Rail}_{Photodiode}}
\end{equation}
$F_1$ is calculated from the dark hall measurements with the PMT; it has
systematic effects due to measurements of the distance between PMT and the
drum or rail source; reflections and stray light; misaligned pointing of the
rail light source; and PMT stability. $F_2$ is calculated based on the PMT
quantum efficiency and the source spectra of the drum and rail light source; if
the PMT QE is not flat over the two spectra and the two source spectra are not
identical, then a systematic effect is introduced.  $F_3$ is calculated from the
dark box measurements using the calibrated photodiode; it may have systematic effects due
to photodiode calibration; reflections in the dark box; and distance
measurements from the rail light source to the photodiode. $F_4$ is calculated
from the NIST--provided calibration of the photodiode and the area of the
photodiode mask, and our measured emission spectrum for the rail light source.
   
A typical value calculated from a calibration curve taken in 2010 gives radiant
drum intensity
$I^{Drum}_0~=~5.70~\pm~0.12~\times10^8~\left[\frac{photons}{sr \cdot pulse}\right]$ 

The possible sources of systematic error mentioned above are discussed in the next section.

\section{Systematic checks and uncertainties}
\label{sec:syst_Uncertain}

The distance between the drum or the rail light source and the PMT is known to a
centimeter at worst; this uncertainty enters in the square of the distance.
For the drum at 15.5 m this uncertainty of 0.01 m results in a 0.1\% systematic
on $I^{Drum}_0$.  Similarly, the closest distance of the rail light source to
the PMT (10 m) results in a conservative 0.2\% uncertainty on $I^{Drum}_0$.  In
the dark box we measure the distance from the rail source to the photodiode to
about 0.5 mm, and this results in a 1\% overall uncertainty.

We studied reflections in the dark hall by removing the anti-reflective baffles
and covers on the floor, and we found no measurable change in the PMT response
to the drum or rail source. Evidently the small entrance hole to the dark box
provided effective baffling.  Stray light and reflections could be
indistinguishable from errors in distance measurements; we used a semi-analytic
calculation to study this issue and are able to bound the effect of stray light
to 0.1 \% on $I^{Drum}_0$.

If the rail light source luminous surface were not perpendicular to the optical
axis, say, while on the rail, or had a different angle to the axis in the dark
box, then a systematic effect results.  Mechanical measurements limit this angle
to about 4 degrees, resulting in a systematic uncertainty on the drum intensity
of 0.6\%.

The time required to make a full calibration of the drum, including drum
intensity, rail light source work with the PMT, and dark box measurements with
the calibrated photodiode, is about three hours.  The systematic due to PMT stability was
estimated by making repeated measurements of the drum or rail source on a time
scale of a several hours or on different days.  The repeatability of the system
(dominated by the PMT) is at the 0.4\% level.  The stability of the LED light
source itself was discussed above, and is at the 0.1\% level over a few days.

NIST provides an uncertainty on calibration of the photodiode (0.8\%) and on
the uncertainty in the area of the mask for the photodiode (0.1\%).

To estimate the deviation of $F_2$ from unity, we measured, as a function of
wavelength, the reflectivity of the Tyvek in the drum and the Teflon face of the
drum and the transmission of the Teflon face.  Convoluting these factors,
allowing for several reflections in the drum before transmission through the
face, results in an estimate of the drum emission spectrum.  In the same way we
calculated the spectrum of the rail light source.  We convoluted these spectra
against the PMT quantum efficiency, and we find $F_2 = 1.001$.  We do not make
this small correction, rather we quote a 0.1\% systematic uncertainty.

Similarly the convolution of the rail light source spectrum with the calibrated
photodiode is uncertain due to both the uncertainty in the photodiode calibration
and the rail light source spectrum. But in this case, since the absolute
calibration of the NIST-provided calibration appears alone (not in a ratio as in
$F_2$ for the PMT measurements) we quote the full 0.8\% NIST uncertainty.

The calibration of the Keithly electrometer is uncertain at the 1\% level, provided
by the manufacturer.

\subsection{Deviations of measured points from $1/{r^2}$}

Shown in the lower portion of Fig.~\ref{fig:Darkbox_Residual} are the
deviations of measurements taken at all distances with the $1/{r^2}$ fit to the
NIST points; the points from the rail source with the PMT are laid on the curve
as discussed above. Deviations of the points from $1/{r^2}$ are all below 1\%,
mostly within the statistical uncertainties, and we observe no significant pull
of the fit.  The geometrical configurations of the dark hall and the dark box
are potential sources of systematic uncertainties due to reflections from walls
or blockage of direct line of sight from detectors to all points on the light
source.  We take the lack of a systematic deviation of points from $1/{r^2}$ in
Fig.~\ref{fig:Darkbox_Residual} as confirmation that such effects are quite
small.

\subsection{Uncertainties}

Table~\ref{tab:uncertainties} lists the uncertainties associated with the
absolute intensity calibration of the drum light source.  Statistical
uncertainties in the terms $<Q>$, $<H_{LS}>$ and $<H_D>$ 
from Eqn.~\ref{eq:drum_intensity_nofph} are vanishingly small and are not
listed.  Repeated measurements of these terms are dominated by the drift in
light source intensity and detector response, which are listed. The overall
uncertainty in the calibration of the Auger Observatory Fluorescence Detectors
involves additional factors such as temperature variations, reflections in the
FD apertures and FD data analysis, which are not discussed here.

\begin{table}[h]
\begin{center}
\caption{Uncertainties for the drum intensity absolute calibration.}
\label{tab:uncertainties}
\begin{tabular}{llr}
\hline
\hline
                    &  Quantity  &  Uncertainty (\%)        \\
\hline
\quad Rail light source: &Stability                  &        0.1    \\

		    &Rail Light Source and Drum relative emission spectra	&	0.1	\\
		    & Rail light source alignment   &   0.6 \\

\quad Distances:     &Photodiode to rail light source &   1.0   \\
\quad               &PMT to rail light source       &   0.2   \\
                    &PMT to drum face               &   0.1   \\
\quad Stray light:  &		      &   0.1 \\

\quad Photodiode:   &NIST calibration @365 nm  	    &   0.8    \\
                    &Active area                    &   0.1    \\
                    &Electrometer calibration       &   1.0    \\
                    &Electrometer configuration repeatability &        1.0    \\

\quad PMT :  	&Stability         		    &   0.4   \\   
\hline      
Total uncertainty    &                               &   2.1    \\
\hline      
\hline      

\end{tabular}
\vspace{0.5cm}
\end{center}
\end{table}

\section{Conclusions}
\label{sec-conclusions}

The measurement in the laboratory of the absolute intensity of the 2.5 m
diameter drum calibration light source is the first step in the production
calibration of the Auger Observatory fluorescence detectors.  The sensitivity of
the NIST-calibrated photodiode used as a calibration reference is some 4 orders
of magnitude below that required to measure the drum intensity directly,
forcing us to use  some method of boosting the output to a measurable level for
calibration.  Working in a 17~m long dark hall allowing measurements at varying
distances between the light source and detectors, we have established a simple
procedure using the $1/{r^2}$ reduction of light flux, eliminating use of
neutral density filters on an optical bench as used previously for this
reduction.  The use of this technique has significantly reduced the dominant
uncertainties from those related to use of the neutral density filters, leading
to an overall uncertainty in the light source intensity of 2.1\%.

\acknowledgments

We acknowledge the contributions (and frustrations) of earlier collaborators in
this Auger Observatory FD calibration effort, especially G. Hofman,
R. Meyhandan, R. Knapik, and P. Bauleo.

\end{document}